\documentclass[pdflatex,sn-mathphys]{sn-jnl}

\jyear{2021}%

\theoremstyle{thmstyleone}%
\theoremstyle{thmstyletwo}%

\theoremstyle{thmstylethree}%

\usepackage{graphicx}
\graphicspath{ {./images/} }
\usepackage{amsmath}
\usepackage{multirow}
\usepackage{comment}
\usepackage{enumitem}

\usepackage{caption} 
\usepackage{subcaption}
\usepackage{lineno}

\begin{document}

\title[]{iDDS: Intelligent Distributed Dispatch and Scheduling for Workflow Orchestration}


\author*[1]{\fnm{Wen} \sur{Guan}}
\author*[1]{\fnm{Tadashi} \sur{Maeno}}
\author[2]{\fnm{Aleksandr} \sur{Alekseev}}
\author[2]{\fnm{Fernando Harald} \sur{Barreiro Megino}}
\author[2]{\fnm{Kaushik} \sur{De}}
\author[1]{\fnm{Edward} \sur{Karavakis}}
\author[1]{\fnm{Alexei} \sur{Klimentov}}
\author[3]{\fnm{Tatiana} \sur{Korchuganova}}
\author[2]{\fnm{FaHui} \sur{Lin}}
\author[1]{\fnm{Paul} \sur{Nilsson}}
\author[1]{\fnm{Torre} \sur{Wenaus}}
\author[1]{\fnm{Zhaoyu} \sur{Yang}}
\author[1]{\fnm{Xin} \sur{Zhao}}

\affil[1]{\orgname{Brookhaven National Laboratory}, \city{Upton}, \state{NY}, \country{USA}}

\affil[2]{\orgname{University of Texas at Arlington}, \city{Arlington}, \state{TX}, \country{USA}}

\affil[3]{\orgname{University of Pittsburgh}, \city{Pittsburgh}, \state{PA}, \country{USA}}




\abstract {
The \textit{intelligent Distributed Dispatch and Scheduling (iDDS)} service is a versatile workflow orchestration system designed for large-scale, distributed scientific computing. iDDS extends traditional workload and data management by integrating data-aware execution, conditional logic, and programmable workflows, enabling automation of complex and dynamic processing pipelines. Originally developed for the ATLAS experiment at the Large Hadron Collider, iDDS has evolved into an experiment-agnostic platform that supports both template-driven workflows and a Function-as-a-Task model for Python-based orchestration.

This paper presents the architecture and core components of iDDS, highlighting its scalability, modular message-driven design, and integration with systems such as PanDA and Rucio. We demonstrate its versatility through real-world use cases: fine-grained tape resource optimization for ATLAS, orchestration of large Directed Acyclic Graph (DAG) workflows for the Rubin Observatory, distributed hyperparameter optimization for machine learning applications, active learning for physics analyses, and AI-assisted detector design at the Electron–Ion Collider.

By unifying workload scheduling, data movement, and adaptive decision-making, iDDS reduces operational overhead and enables reproducible, high-throughput workflows across heterogeneous infrastructures. We conclude with current challenges and future directions, including interactive, cloud-native, and serverless workflow support.
}

\maketitle

\section{Introduction}\label{sec1}

The growing complexity of scientific computing poses major challenges for workflow orchestration. Modern large-scale experiments in high-energy physics, astronomy, and related domains routinely combine heterogeneous tasks—data management, simulation, machine learning, and analysis—executed across geographically distributed infrastructures. Traditional workflow and workload management systems provide essential scheduling and data handling capabilities, but they are often limited in their ability to express dynamic dependencies, integrate data availability directly into execution logic, or support iterative and adaptive workloads such as hyperparameter optimization and active learning.

The \textit{intelligent Distributed Dispatch and Scheduling (iDDS)} service was designed to address these gaps. iDDS provides a unified orchestration framework that combines data awareness, conditional execution, and flexible workflow representation to enable fine-grained automation at scale. Unlike conventional workload managers, iDDS treats workflows as programmable objects, supporting both template-based orchestration for well-structured pipelines and a Function-as-a-Task model that allows users to express workflows directly in Python. This dual approach makes iDDS suitable for both production-grade data processing campaigns and rapidly evolving machine learning pipelines.

The key capability of iDDS is to coordinate diverse tasks and activities, reducing operational overhead and increasing automation to improve efficiency. Its main features include: (1) integrating fine-grained data availability and movement into workflow logic for data-aware orchestration, (2) supporting complex workflow management such as Directed Acyclic Graphs (DAGs), conditional branching, and polymorphic workflows, (3) enabling iterative execution with parameter sweeps, iterative sequences, and distributed hyperparameter optimization, (4) providing a code-driven design that allows workflows to be expressed programmatically using Python functions, and (5) integrating with large-scale, geographically distributed workload management systems to ensure scalability.

Originally developed for the ATLAS~\cite{atlas} experiment at the Large Hadron Collider (LHC)~\cite{lhc}, iDDS has since evolved into a general-purpose orchestration platform adopted by multiple projects, including the Rubin Observatory~\cite{rubin} and the Electron–Ion Collider (EIC)~\cite{eic}. It has been successfully integrated with the Production and Distributed Analysis (PanDA)~\cite{panda} system for workload management and with Rucio~\cite{rucio} for data management. PanDA handles the scheduling of workloads across large-scale, heterogeneous distributed computing resources, while Rucio manages data movement among collaborating institutions. Its applications range from large-scale data reprocessing on tape, to management of complex DAG workflows, to distributed hyperparameter optimization and AI-assisted detector design.

In this paper, we present the concepts, architecture, and implementation of iDDS, and illustrate its versatility through real-world use cases. We conclude with a discussion of achievements, challenges, and future directions.

\section{Concepts}\label{sec:concept}

\subsection{Core Concepts}
\label{subsec:core_concepts}

The iDDS system is built around four fundamental concepts that collectively define the structure and logic of how workflows are represented, managed, and executed.

\paragraph{Work.}
A \textit{Work} unit is the atomic executable entity within a workflow. Each Work unit encapsulates a self-contained task  (such as data transformation, inference, simulation, and filtering) and carries metadata describing its execution state, dependencies, inputs, and outputs. Each task consists of a group of jobs with similar attributes, which serve as the actual units of execution. Work units can be run independently or composed into a larger workflow, with their progress and state tracked throughout their lifecycle.

\paragraph{Workflow.}
A \textit{Workflow} is a collection of Work units connected through well-defined dependency relationships. This is represented as a Directed Acyclic Graph (DAG) that encodes the sequence and logic of execution, including parallelization, ordering, and data transfer. Workflows can be specified statically at submission time or dynamically expanded in response to runtime conditions. The DAG-based representation enables users to express complex processing logic while allowing iDDS to optimize overall execution strategies.

\paragraph{Condition.}
A \textit{Condition} is a control structure that guides the execution of a workflow by evaluating runtime information, such as the output of previous \textit{Work} units or system metrics. Based on this evaluation, it determines whether and how subsequent Work units are executed. Conditions allow for branching, delays, failure handling, and adaptive behavior within workflows.

\paragraph{Parameter.}
\textit{Parameters} are key-value pairs that are passed into Work units and Workflows to influence their execution behavior. They may define runtime settings, dataset identifiers, model configurations, or execution thresholds. Parameters can be hierarchical and dynamically generated during workflow execution, supporting advanced techniques such as hyperparameter search or data-driven configuration.

\subsection{Workflow Representation Styles}
\label{subsec:orchestration_styles}

To support a wide variety of use cases, iDDS provides two primary styles of workflow representation, each addressing different user requirements and levels of abstraction. These styles can be used independently or combined, providing users with flexibility to balance between abstraction and control in their workflow design.

\paragraph{Template-based Representation.}
In this style, users define workflows declaratively using templates. These templates describe the structure and dependencies among Work units, along with the associated Conditions and Parameters. While templates can be dynamically generated, they become static or semi-static once submitted to iDDS. This approach is particularly suited for well-defined workflows such as data production chains, where the workflow logic remains constant but the parameters (e.g., input datasets) vary at runtime. Template-based representation offers high reusability and can be validated prior to execution.

\paragraph{Code-based Representation (a.k.a Function-as-a-Task).}
For more dynamic and logic-heavy workflows, iDDS supports a code-driven style where users define their workflows programmatically using Python functions. Each function acts as a Task, and its outputs can be analyzed at runtime to determine subsequent execution steps. This style, inspired by the Function-as-a-Service~\cite{funcx} paradigm, allows for complex runtime decisions, loop constructs, and integration with external APIs. It is particularly well-suited for use cases like machine learning pipelines or real-time analysis, where execution logic must adapt based on intermediate results.

\section{System Architecture}
\label{sec:architecture}

This section describes how the core concepts of iDDS are implemented in practice, including the system’s modular design, data flow, and key components responsible for orchestration, communication, and execution.

\begin{figure}[h]
\centering
\includegraphics[width=10cm]{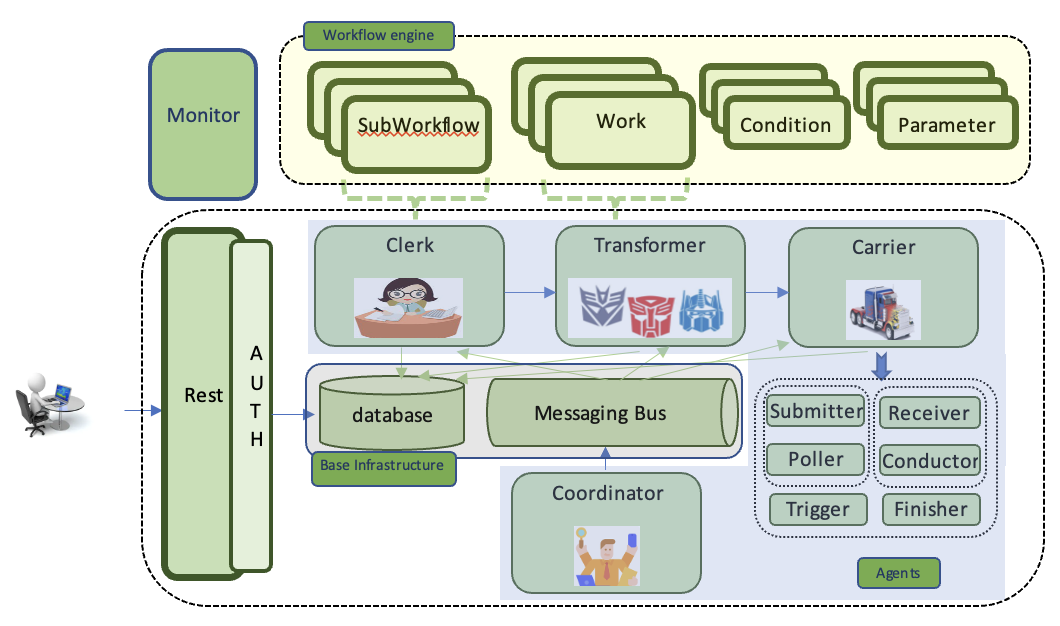}
\caption{Schematic overview of the iDDS architecture, illustrating its main components: (1) Workflow engine, (2) base infrastructure, (3) RESTful service, (4) Agents, and (5) Monitors.}
\label{fig:idds_schematic_view}
\end{figure}

Fig.~\ref{fig:idds_schematic_view} provides a schematic overview of the iDDS architecture, which comprises five main components: the workflow engine, base infrastructure, RESTful service, agents, and monitors.

\subsection{Workflow Engine}
\label{subsec:workflow_engine}

The workflow engine implements the core abstractions defined in Section~\ref{subsec:core_concepts}—\emph{Work}, \emph{Workflow}, \emph{Condition}, and \emph{Parameter}—as concrete classes. Each class encapsulates a \emph{Template} (static logic) and a \emph{Metadata} object (dynamic runtime context), together forming the foundation for reusable and adaptive workflows. 

\begin{itemize}
    \item \textbf{Templates:} define reusable workflow blueprints with partially fixed structures and parameters.  
    \item \textbf{Metadata:} capture dynamic runtime information, enabling workflows to evolve adaptively based on execution context.  
\end{itemize}

These objects are persisted in the database, executed by agents, and tracked throughout their lifecycle.

\subsubsection{Directed Acyclic Graph (DAG)}\label{subsubsec:dag}

iDDS supports both Directed Acyclic Graph (DAG) and cyclic graph structures at the task and job levels, integrating seamlessly with workload management systems such as PanDA to orchestrate large-scale processing workloads.

At the task level, iDDS implements a Directed Graph (DG) engine that manages acyclic and cyclic dependencies. \textit{Templates} define DAG- and loop-based workflows, while \textit{Metadata} and custom conditions control branching and execution logic. This enables dynamic, adaptive execution paths based on runtime status.

At the job level, DAG support manages fine-grained job dependencies. iDDS automatically evaluates these relationships and incrementally releases downstream jobs as their dependencies are satisfied.

\subsubsection{Workflow Execution and Tracking}\label{subsubsec:workflow_execution_and_tracking}

Each \textit{Work} unit and \textit{Workflow} is represented in the database with attributes such as status, timestamps, dependencies, input/output datasets, parameter bindings, and \textit{Metadata}. iDDS employs a state machine to track the lifecycle of each \textit{Work} unit, from submission through execution to completion or failure. Execution is initiated by agent components (see Section~\ref{subsec:agents}) and monitored through a combination of periodic polling and event-driven updates.

For execution, each \textit{Work} unit is serialized and submitted to workload management systems such as PanDA or HTCondor~\cite{htcondor}, where it runs as a distributed job. On the compute node, a lightweight wrapper reconstructs the \textit{Work} object and executes its logic. iDDS agents then monitor job progress, evaluate \textit{Conditions}, propagate and bind \textit{Parameters}, and track results to ensure correct orchestration across the workflow.

\subsubsection{Function-as-a-Task}\label{subsubsec:function_as_a_work}

\begin{figure}[h]
\centering
\includegraphics[width=10cm]{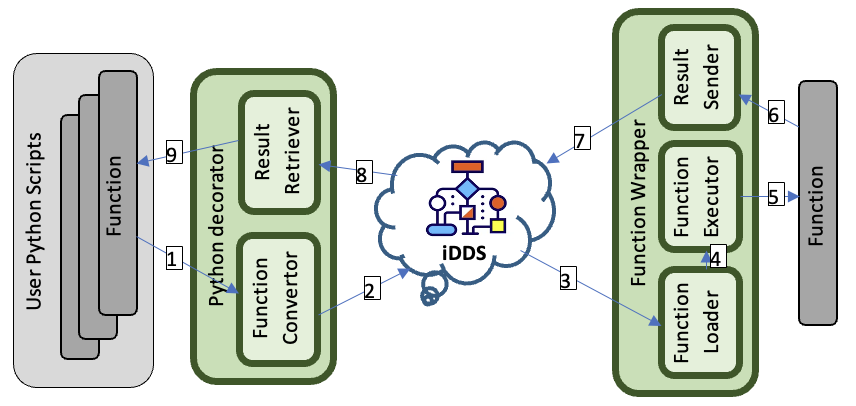}
\caption{Function-as-a-Task workflow: (1) Local Python functions are serialized and converted into executable \textit{Work} units using Python decorators; (2) iDDS submits the \textit{Work} units as \textit{Tasks} to remote workers via a workload system; (3) A function wrapper loads the \textit{Work} units and executes the function, collects the results, and sends them back to iDDS; (4) The local Python decorator retrieves and returns the results to the function caller.}
\label{fig:function_as_a_work}
\end{figure}

The core idea of Function-as-a-Task is to transparently convert functions into \textit{Work} objects using Python decorators, which are then submitted as \textit{Tasks} to remote workers via a workload management system. The user-defined Python script serves as the \textit{Workflow} that controls execution logic (Fig.~\ref{fig:function_as_a_work}). \textit{Workflow} scripts can run locally or on iDDS-managed clusters. For security, iDDS executes them on a sandboxed HTCondor cluster, isolating untrusted code from the main server infrastructure. This model simplifies the construction of complex workflows—such as machine learning pipelines—by allowing users to express intricate logic and conditional behavior directly in Python.

The \textit{Workflow} unit can be executed either on the client side or on the iDDS server side:

\begin{itemize}
  \item \textbf{Client side}:  
  The \textit{Workflow} may be embedded in a user's main script (e.g., a machine learning script) and executed directly on the client side. If the user submits the main script to PanDA as a PanDA job, the \textit{Workflow} is executed as part of that job.

  \item \textbf{iDDS server side}:  
  Lightweight workflows can be executed on the iDDS server to reduce latency and improve responsiveness. The server hosts a small HTCondor cluster dedicated to running \textit{Workflow} scripts. HTCondor ensures script isolation and prevents excessive concurrent workflow executions that could overload the system.
\end{itemize}

\textit{Work} unit execution proceeds in two stages:

\begin{itemize}
  \item \textbf{Serialization and distribution:} Annotated functions are serialized and wrapped as \textit{Work} objects, allowing transparent submission to workload management systems as Tasks. During this stage, the source code and execution environment are packaged into a ZIP archive and uploaded to a centrally managed HTTP cache, where access is protected through x509/OIDC-based authorization.
  The \textit{Work} objects are then submitted as tasks or jobs, with hooks tracking their execution status and results.
  \item \textbf{Execution:} The \textit{Work} object is deployed across distributed resources through workload management systems, together with its source code and execution environment retrieved from the HTTP cache. An enhanced wrapper reconstructs the \textit{Work} object and executes the function, while results are returned asynchronously via STOMP~\cite{stomp} or RESTful HTTP.
\end{itemize}

This design enables seamless, scalable execution of complex Python functions on distributed infrastructures such as PanDA, requiring no significant code adaptation while ensuring reliable and responsive result retrieval.

\subsection{Base Infrastructure: Database and Event Bus}
\label{subsec:core_infrastructure}

\begin{figure}[h]
\centering
\includegraphics[width=10cm]{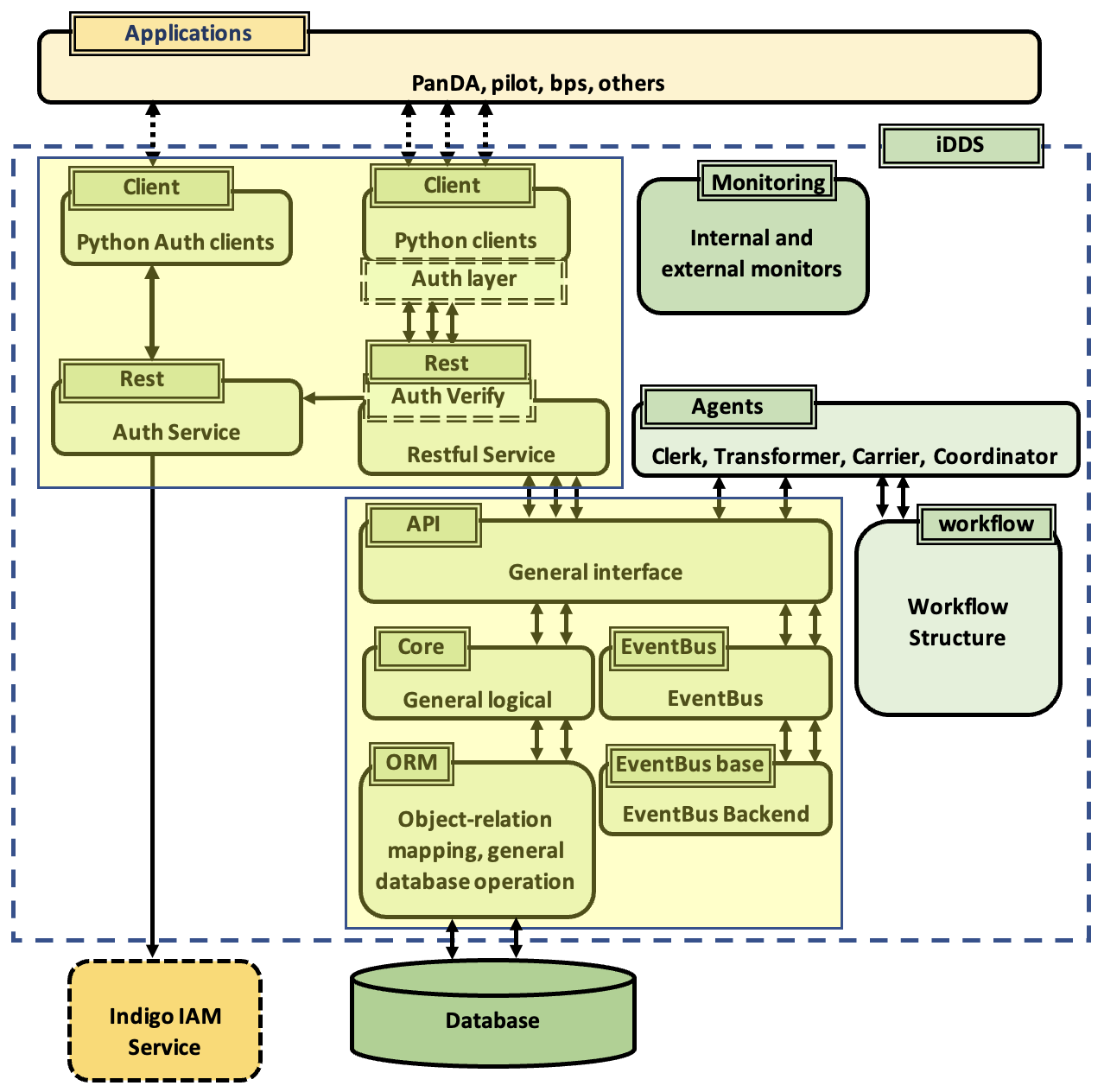}
\caption{iDDS architecture.}
\label{fig:idds_architecture}       
\end{figure}

The database and the event bus are the backbone of iDDS, serving as the base infrastructure upon which all other system components are built, as shown in Fig.~\ref{fig:idds_architecture}. 

\subsubsection{Database}\label{subsubsec:database}

iDDS employs a relational database as its central repository for persisting workflow states, work unit dependencies, data collections, and scheduling metadata. The schema incorporates versioning to support both backward compatibility and forward extensibility.

The database serves two critical roles. First, it records user-submitted workflow requests and captures the relationships among workflow objects, ensuring persistence and traceability. Second,  it maintains the status of workflow objects for organizing and coordinating system operations. This structured status tracking enables iDDS agents to efficiently identify and operate on tasks that are ready for execution, ensuring reliable and scalable workflow management.

iDDS leverages SQLAlchemy~\cite{sqlalchemy} for object–relational mapping, enabling automatic mapping of Python objects to relational tables. SQLAlchemy supports dynamic schema creation and teardown, simplifying testing and deployment, while its compatibility with multiple backends—such as Oracle, PostgreSQL, MySQL, and SQLite—allows iDDS to operate across diverse platforms and migrate between databases with minimal effort. Database schema versioning is managed by Alembic~\cite{alembic}, which automates upgrades and downgrades to ensure the schema evolves consistently with the codebase. This approach streamlines maintenance, supports agile development, and preserves data integrity across deployments.

\subsubsection{Event Bus}\label{subsubsec:event_bus}

iDDS employs an event bus based on the publish-subscribe model to enable asynchronous, decoupled communication among system components. This central communication infrastructure allows event producers and consumers to interact without being directly linked.
Events—such as task completions, data availability, error signals, and status updates—are published to the bus and consumed by agents or orchestrators to trigger subsequent execution stages. This design enhances responsiveness and modularity, allowing agents to react immediately to workflow state changes and significantly accelerating system operations.

iDDS implements multiple event bus backends to support various deployment needs (external messaging services like ActiveMQ~\cite{activemq} can also be integrated as backends):
\begin{itemize}
\item \textbf{LocalEventBus:} A lightweight implementation based on a Python dictionary, enabling fast in-process event delivery. It is suitable for single-process deployments but not usable when multiple iDDS server processes are active.
\item \textbf{DBEventBus:} A database-backed event bus that stores events persistently, enabling distributed delivery across agents on different hosts. Performance depends on the underlying database system.
\item \textbf{MsgEventBus:} A high-throughput, distributed event bus built on the ZeroMQ~\cite{zeromq} messaging library. While efficient, it requires application-level logic to handle message routing and delivery guarantees.
\end{itemize}

\subsection{RESTful Service}
\label{subsec:restful_service}

The RESTful service offers a standardized interface for both users and external systems, exposing endpoints for workflow submission, status queries, data caching, catalog access, and more. Designed according to REST principles, these APIs ensure ease of integration and broad compatibility with external platforms.

\begin{figure}[h]
\centering
\includegraphics[width=10cm]{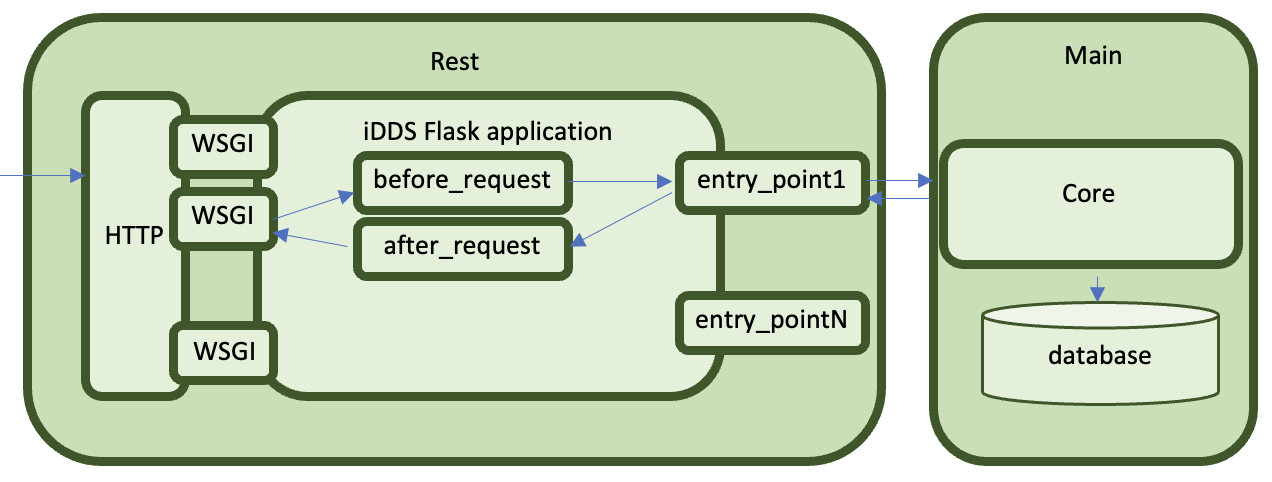}
\caption{iDDS RESTful service architecture: (1) the iDDS application is implemented using Flask and served via WSGI daemons behind an Apache HTTP server; (2) authentication and authorization are enforced through filters applied in the \textit{before\_request} hook; (3) validated requests are dispatched to corresponding entry points.}
\label{fig:idds_rest}
\end{figure}

The service is implemented as a Flask~\cite{flask} application deployed using WSGI~\cite{wsgi} daemons behind an Apache HTTP server, as shown in Fig.~\ref{fig:idds_rest}. The HTTP server passively listens for incoming HTTPS requests and relays them to WSGI daemons, which invoke the Flask application. Within Flask, authentication and authorization filters are applied via the \emph{before\_request} hook. Upon successful authorization, requests are dispatched to the appropriate entry points, which invoke iDDS core functionalities.

\subsubsection{RESTful Entry Points}\label{subsubsec:restful_entry_points}

The RESTful API is organized into multiple logical groups:
\begin{itemize}
    \item \textbf{authentication}: Handles identity flows such as OIDC token generation.
    \item \textbf{ping}: Simple health check endpoint to verify server availability.
    \item \textbf{request}: Submit, update, or query the status of workflow requests.
    \item \textbf{cache}: Upload user-defined code or payloads to the iDDS server.
    \item \textbf{catalog}: Query data availability and status for workflow-related datasets.
    \item \textbf{monitor}: Retrieve monitoring information for workflows and their associated tasks.
    \item \textbf{message}: Send control messages (e.g., abort workflow).
    \item \textbf{log}: Access logs for workflows and work items.
\end{itemize}

This modular organization ensures a scalable and user-friendly interface for interacting with iDDS.

\subsubsection{Authentication and authorization}\label{subsubsec:authentication_and_authorization}

iDDS supports both OpenID Connect (OIDC) tokens and X.509 certificates for flexible, secure authentication and authorization.

\paragraph{OIDC-Based Authentication and Authorization.}
OIDC is supported via integration with Indigo IAM~\cite{iam}, enabling identity federation through providers like Google, CILogon~\cite{cilogon}, or dex~\cite{dex}. The flow (illustrated in Fig.~\ref{fig:idds_auth}) consists of three stages:

\begin{figure}[h]
\centering
\includegraphics[width=10cm]{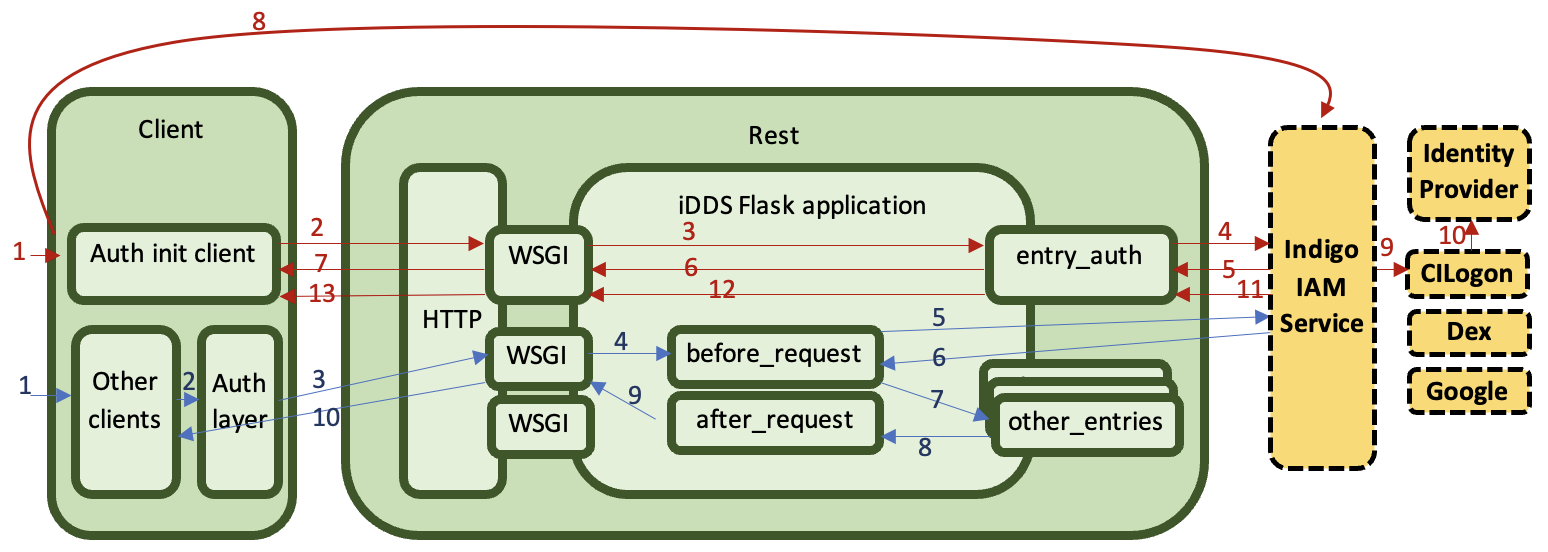}
\caption{iDDS OIDC-based IAM: (1) authentication stage (red); (2) authorization stage (blue).}
\label{fig:idds_auth}
\end{figure}

\begin{itemize}
    \item \textbf{Registration}: Users register with Indigo IAM through a trusted identity provider. During registration, group memberships are assigned to control access rights.
    \item \textbf{Authentication}: The user initiates the flow by contacting the authentication entry point. The system responds with a login URL. The user authenticates through the identity provider, and a token containing identity and group information is returned.
    \item \textbf{Authorization}: When users access protected resources, their token is parsed by the authorization logic in Flask's \emph{before\_request}. The token is validated against Indigo IAM, and access permissions are determined based on group roles embedded in the token. The resolved roles and permissions are then cached for some time to optimize later access.
\end{itemize}

\paragraph{X.509-Based Authentication and Authorization.}
For legacy and grid-based environments, iDDS also supports X.509 certificates. The Apache HTTP server integrates with the GridSite library~\cite{gridsite} to validate user certificates and enforce access control based on certificate attributes.

\subsection{Agents}
\label{subsec:agents}

Agents are stateless, autonomous components responsible for executing and coordinating iDDS workflows. Each agent specializes in a specific role and interacts with the central database and event bus to receive tasks, report progress, and trigger downstream operations. Agents are horizontally scalable and operate asynchronously to support high-throughput processing.

\subsubsection{Agent Roles}\label{subsubsec:agent_roles}

Workflows are initiated via the RESTful service and registered in the central database. The agents then process each workflow as follows:

\begin{itemize}
    \item \textbf{Clerk}: Decomposes Workflow by creating and managing \textit{Work} objects based on defined \textit{Conditions} and \textit{Parameters}.
    \item \textbf{Transformer}: Prepares \textit{Work} objects for execution, ensuring all prerequisites (e.g., input data) are met and selecting appropriate execution environments.
    \item \textbf{Carrier}: Manages communication with external workload management systems, including submission of \textit{Work} objects and monitoring their execution status.
    \item \textbf{Coordinator}: Optimizes event delivery within the event bus by aggregating and prioritizing messages to prevent bottlenecks.
\end{itemize}


\subsubsection{Agent Details}\label{subsubsec:agent_details}
\paragraph{Clerk}
\label{clerk}
The Clerk agent decomposes Workflow and generates Work objects. During workflow execution, it evaluates \textit{Condition} objects to determine if new \textit{Work} objects should be created or if the workflow should terminate. When a new \textit{Work} object is needed, the Clerk references \textit{Parameter} objects to generate inputs. It continuously monitors the state of active \textit{Work} objects and, upon their completion, updates related \textit{Condition} and \textit{Parameter} objects to drive downstream execution.

\paragraph{Transformer}
\label{transformer}
The Transformer agent coordinates the execution of \textit{Work} objects. It verifies that all execution prerequisites—such as input data—are met and selects the appropriate workload system based on availability, efficiency, and policy constraints. The Transformer ensures that each \textit{Work} object is executed in an optimal environment, improving resource utilization and minimizing delays.

\paragraph{Carrier}
\label{carrier}
The Carrier agent interfaces with external workload management systems to handle the submission and tracking of the \textit{Work} execution. It uses a message-based channel to exchange status and control information efficiently. The Carrier comprises several sub-agents, each performing specific tasks:
\begin{itemize}
    \item \textbf{Submitter}: Submits \textit{Work} objects to the workload management system and returns tracking metadata.
    \item \textbf{Poller}: Monitors execution status of submitted \textit{Work} objects.
    \item \textbf{Finisher}: Finalizes \textit{Work} objects upon completion or failure.
    \item \textbf{Conductor}: Sends execution status updates to external systems.
    \item \textbf{Receiver}: Consumes status messages from the workload system and updates internal records.
    \item \textbf{Trigger}: Evaluates dependency graphs and triggers downstream \textit{Work} objects when conditions are satisfied.
\end{itemize}

\paragraph{Coordinator}
\label{coordinator}
The Coordinator agent enhances the efficiency of the event bus, which enables asynchronous communication across the iDDS system. In large-scale workloads, the event bus may become congested due to high message volumes (e.g., thousands of parallel job updates). The Coordinator addresses this challenge by:
\begin{itemize}
    \item \textbf{Merging Events}: Consolidates similar or redundant messages to avoid unnecessary overhead.
    \item \textbf{Priority Management}: Assigns higher priority to critical operations (e.g., \textit{Work} objects completion) over lower-priority updates.
\end{itemize}
By regulating event flow, the Coordinator ensures agents remain responsive and system operations are not delayed by excessive message traffic.
\subsubsection{Operation Scheduling}\label{subsubsec:operation_scheduling}

Operations in iDDS are fully distributed, with no central scheduler. Specialized agents manage different types of operations and periodically poll the database to identify tasks that have remained idle beyond a configured threshold. Upon detection, they initiate the corresponding actions.

To enhance responsiveness, iDDS integrates an optional event bus that supports asynchronous, event-driven communication. When an agent completes an operation, it emits an event to the bus, triggering other agents responsible for the next steps. This reduces latency and enables prompt reactions to state changes.

To prevent duplicate execution, agents update the status and timestamp of tasks upon triggering, ensuring they are not reprocessed by other agents during polling. While event triggering is the primary mechanism for rapid response, some events may be lost due to network or system issues. To address this, database polling operates in a fallback or "lazy" mode, ensuring that missed events are eventually handled. This hybrid design allows iDDS to balance efficiency and reliability, with the flexibility to disable the event bus when not required.

\subsection{PanDA Integration}
\label{subsec:panda_integration}

\begin{figure}[h]
\centering
\includegraphics[width=11cm]{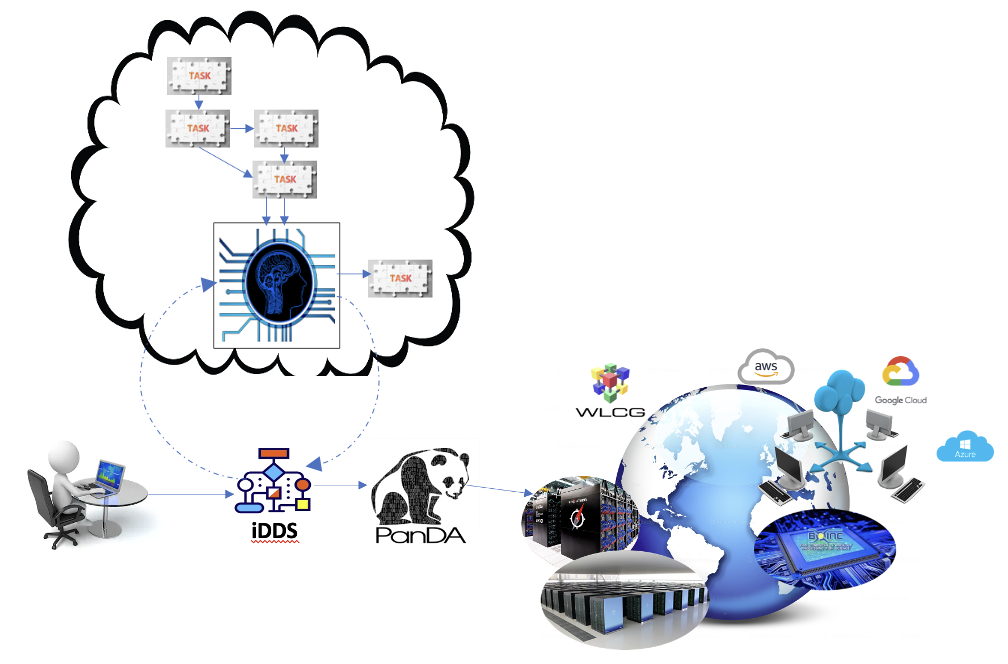}
\caption{An integrated workflow with PanDA and iDDS: iDDS automates complex, dynamic workflows while PanDA schedules workloads across large-scale, distributed, and heterogeneous computing resources.}
\label{fig:fig_ml_panda_idds}
\end{figure}

iDDS is tightly integrated with the PanDA Workload Management System~\cite{panda}, enabling distributed job submission and execution across heterogeneous computing resources within the WLCG and other infrastructures. In this integration, \textit{Work} objects from iDDS workflows are translated into PanDA tasks or jobs, allowing iDDS to handle orchestration while PanDA manages large-scale execution.

PanDA is a robust, production-grade workload management system designed for high-throughput, distributed computing. It excels at managing large-scale and heterogeneous resources across institutions and regions. One of PanDA’s key strengths lies in its abstraction of underlying compute infrastructure, presenting users with a unified interface for job submission and monitoring. This abstraction allows users to deploy complex workloads without needing detailed knowledge of the underlying systems.

By leveraging PanDA, iDDS gains access to transparent, policy-driven scheduling and execution capabilities across a wide array of computing sites. This is particularly beneficial for large-scale machine learning and data processing workflows, where coordination of compute and data placement is critical.

Figure~\ref{fig:fig_ml_panda_idds} illustrates a typical integrated workflow, where iDDS manages the dynamic orchestration of tasks, while PanDA schedules and dispatches jobs to globally distributed resources.

\subsection{Monitoring}
\label{subsec:monitoring}

iDDS provides comprehensive monitoring capabilities to ensure the health, transparency, and efficiency of workflow execution. Monitoring tools are used to track system status, workflow progress, and performance metrics such as throughput, latency, failure rates, and resource usage. Logs from agents and services are collected and visualized in real-time dashboards, with integration support for observability tools like Loki and Grafana to facilitate debugging and performance tuning.

\begin{figure}[h]
\centering
\includegraphics[width=10cm]{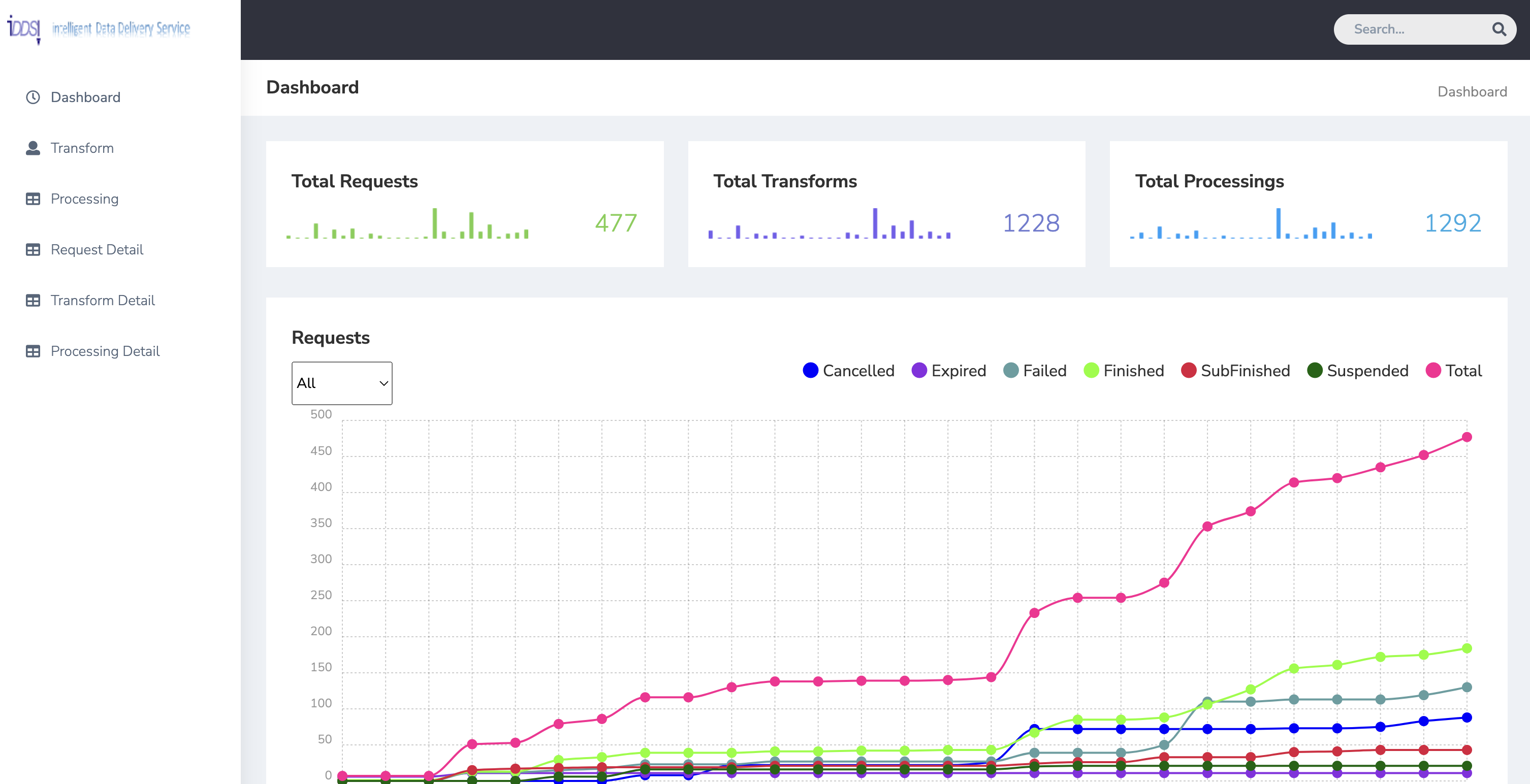}
\caption{Example visualization from the iDDS internal monitor, showing the state of Work and Workflow objects.}
\label{fig:idds_internal_monitor}
\end{figure}

\begin{figure}[h]
\centering
\includegraphics[width=10cm]{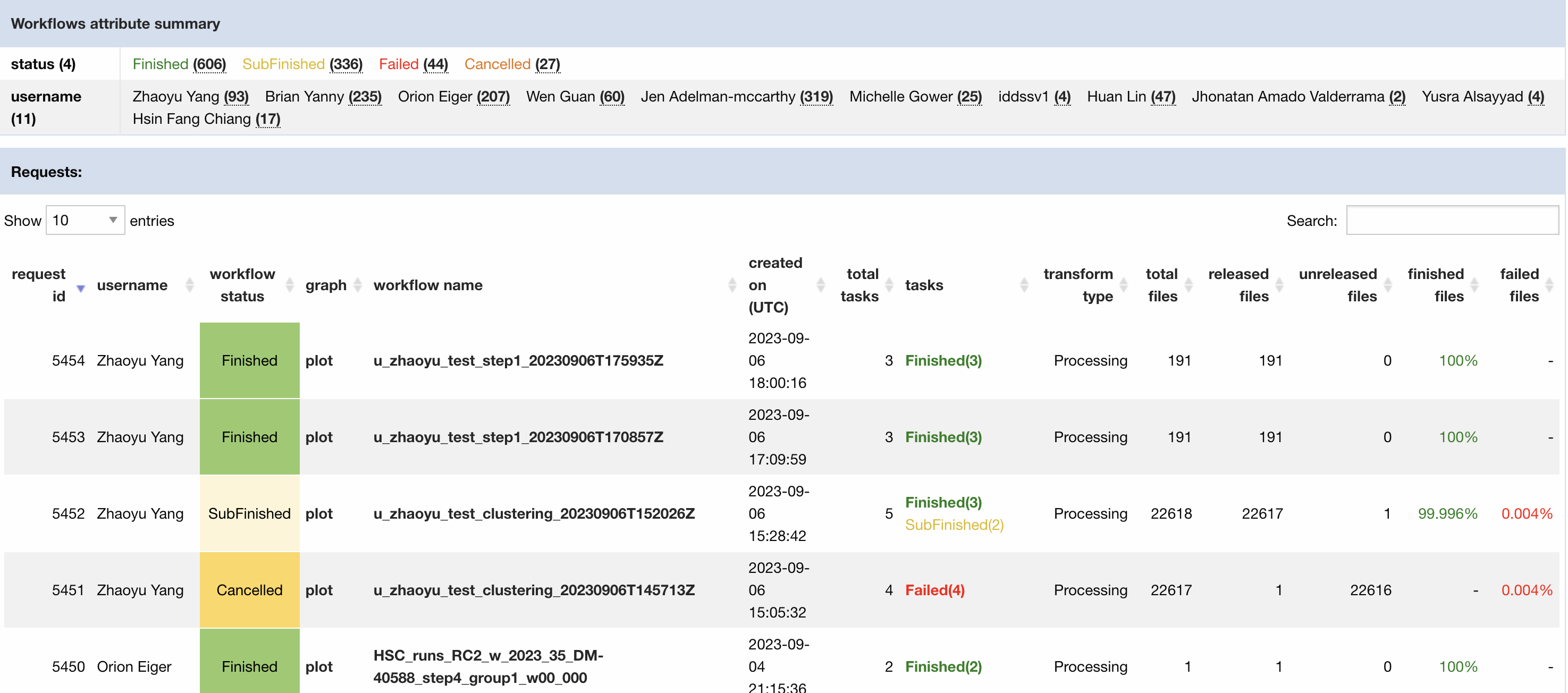}
\caption{Example visualization from the PanDA-integrated monitor, correlating iDDS workflows with PanDA job execution.}
\label{fig:panda_workflow_monitor}
\end{figure}

Internally, iDDS includes a built-in monitoring system that continuously tracks the state of both \textit{Workflow} and \textit{Work} objects. This internal monitor offers real-time visibility into the orchestration lifecycle and supports operational diagnostics, as illustrated in Fig.~\ref{fig:idds_internal_monitor}.

In addition to the native monitor, iDDS integrates with the PanDA monitoring system to correlate workflow-level metadata with job execution status. This integration provides end-to-end visibility into distributed workloads, enabling users to track both high-level workflow orchestration and low-level job execution metrics, as shown in Fig.~\ref{fig:panda_workflow_monitor}.

\section{Use Cases}

iDDS has demonstrated its flexibility and scalability across diverse scientific domains to orchestrate complex workflows, manage large-scale data operations, and facilitate AI-driven scientific discovery. This section highlights use cases that illustrate these capabilities in practice.

\subsection{Optimization of Tape Resource Utilization for ATLAS}
The ATLAS Data Carousel~\cite{datacarousel, datacarousel1} aims to maximize the use of cost-effective tape storage over more expensive disk resources.

\begin{figure}[h]
\centering
\includegraphics[width=10cm]{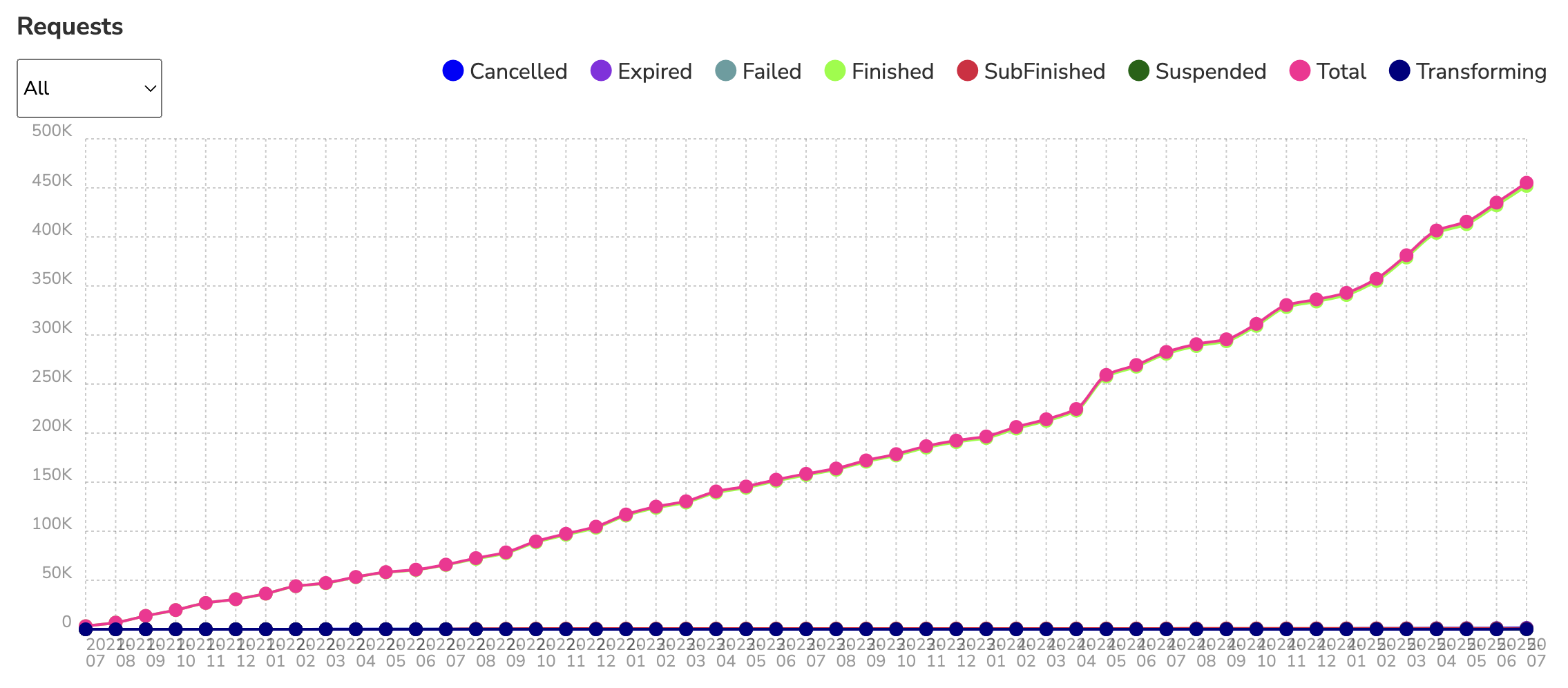}
\caption{Data reprocessing throughput (2021–2025) using the fine-grained Data Carousel with iDDS. The y-axis represents the number of processed requests, each corresponding to a dataset that may span hundreds of gigabytes or more.}
\label{fig:data_carousel1}
\end{figure}

Traditionally the Data Carousel operated at the dataset level due to limitations in the Workflow Management (WFM) and Distributed Data Management (DDM) systems, which resulted in significant overhead and required large disk pools to cache entire datasets before processing could begin. In contrast, iDDS enhances the WFM system with file-level granularity, enabling input data to be processed incrementally as it becomes available from tape~\cite{idds1,idds2}. This fine-grained awareness significantly reduces data staging overhead and eliminates redundant data transfers and caching. This approach allows iDDS to maintain a minimal input data footprint on disk, optimizing the end-to-end resource usage.

Fully integrated into the ATLAS computing infrastructure since mid-2020, iDDS has played a key role in large-scale data reprocessing campaigns. Figure~\ref{fig:data_carousel1} presents the number of processed requests (one per dataset) from 2021 to 2025, illustrating the sustained throughput achieved with the iDDS-enhanced Data Carousel.

\subsection{Complex Workflows for Rubin Observatory}
\label{subsec:rubin_dag}
The Rubin Observatory (LSST) leverages PanDA as both its workflow and workload management system~\cite{panda-rubin}, with iDDS integrated to manage complex task and job dependencies. For each submitted payload, Rubin middleware dynamically generates a workflow graph containing job-level dependencies. These workflows can comprise over 100,000 jobs, forming the vertices of a large DAG.

\begin{figure}[h]
\centering
\includegraphics[width=10cm]{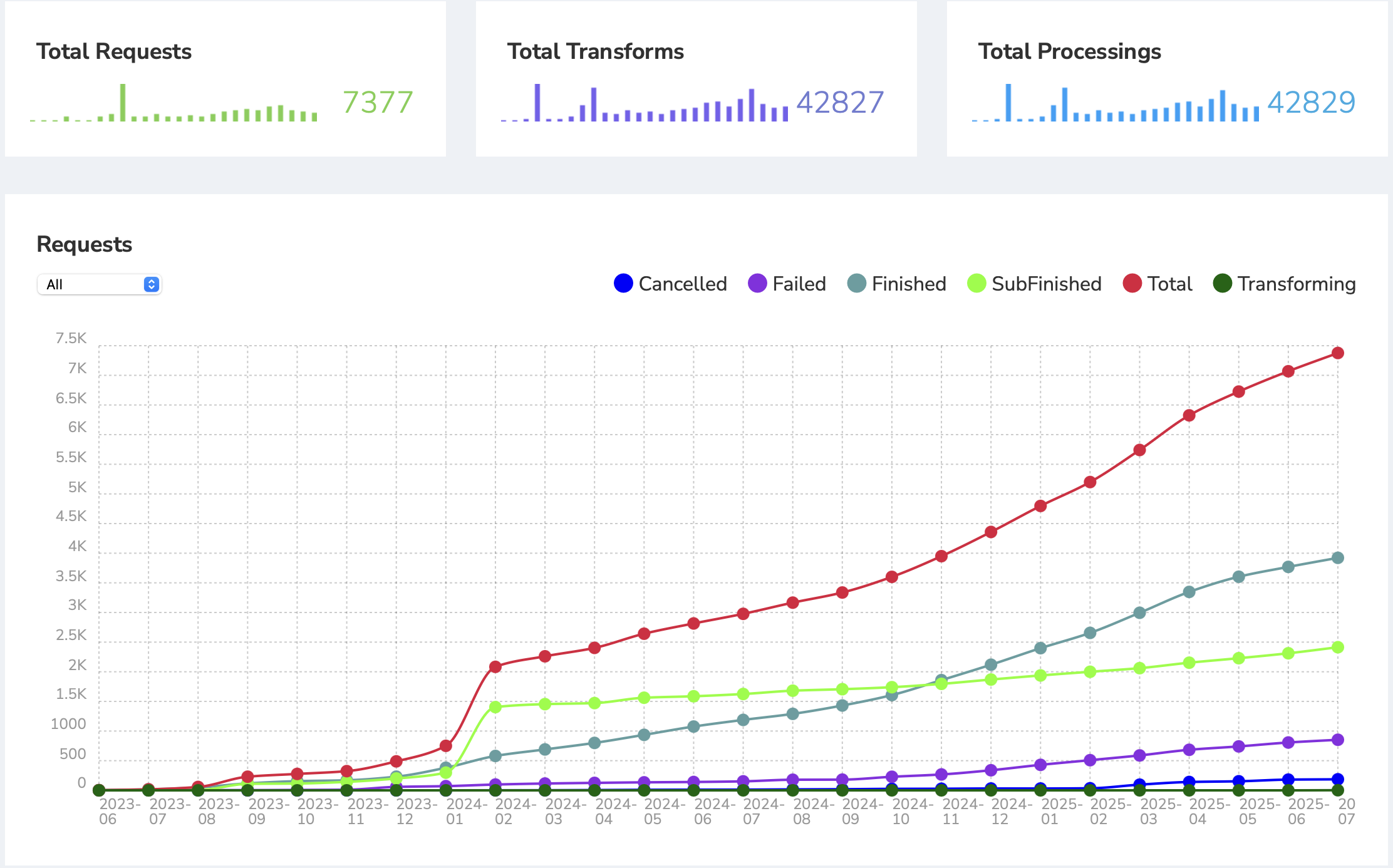}
\caption{Rubin Observatory production activity using the iDDS–PanDA system since late 2021. The monitor displays data from mid-2023 onward, as earlier data have been archived. Each request corresponds to a workflow, and each transform represents a task comprising hundreds of thousands of jobs.}
\label{fig:rubin}
\end{figure}

\begin{figure}[h]
\centering
\includegraphics[width=10cm]{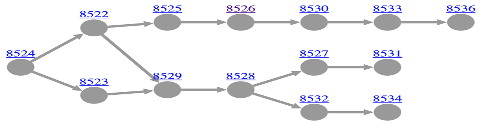}
\caption{Visualization of a DAG showing task-level dependencies within a Rubin Observatory workflow.}
\label{fig:rubin_dag}
\end{figure}

iDDS plays a central role by incrementally releasing jobs based on dependency resolution and messaging triggers. When a job completes, iDDS triggers agents to evaluate dependent jobs and release them as appropriate. At the task level, iDDS manages the execution and release of finalizing steps such as merge tasks.

This system has been in production since mid-2021, with a dedicated PanDA–iDDS instance deployed at SLAC to support Rubin Observatory’s data production campaigns. Over the past few years, it has processed numerous tasks across multiple sites for Rubin ComCam (Commissioning Camera) and DRP (Data Release Production) workflows, as illustrated in Figure~\ref{fig:rubin}. An example task-level DAG from a Rubin workflow is shown in Figure~\ref{fig:rubin_dag}.

\subsection{Distributed Hyperparameter Optimization}
\label{subsec:hpo}

Hyperparameter Optimization (HPO)~\cite{hpo} is a fundamental task in machine learning (ML), aimed at tuning the parameters that govern the training process to achieve optimal model performance. Effective HPO often requires launching and evaluating a large number of training jobs in parallel, making it computationally expensive and technically complex—especially in geographically distributed environments such as grids, HPC systems, and cloud platforms.

iDDS addresses these challenges by providing a fully automated, scalable platform for distributed HPO
~\cite{idds-activelearning}. It seamlessly orchestrates workloads across heterogeneous CPU/GPU resources, and efficiently collects and integrates upstream training results to guide subsequent iterations. iDDS is particularly well-suited for ML workflows involving iterative computation and dynamic decision-making.

\begin{figure}[h]
\centering
    \captionsetup{justification=centering,margin=2cm}
    {\includegraphics[width=5.5cm]{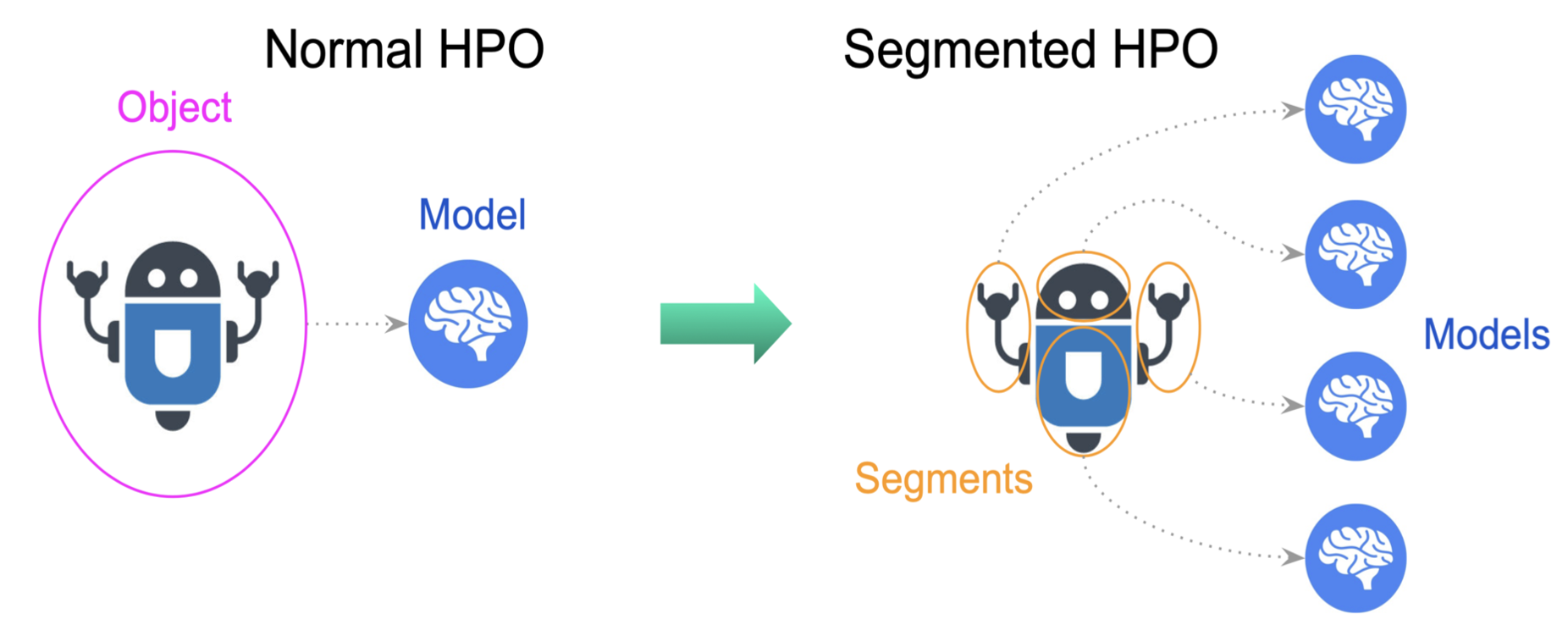} }%
    \qquad
    {\includegraphics[width=5.5cm]{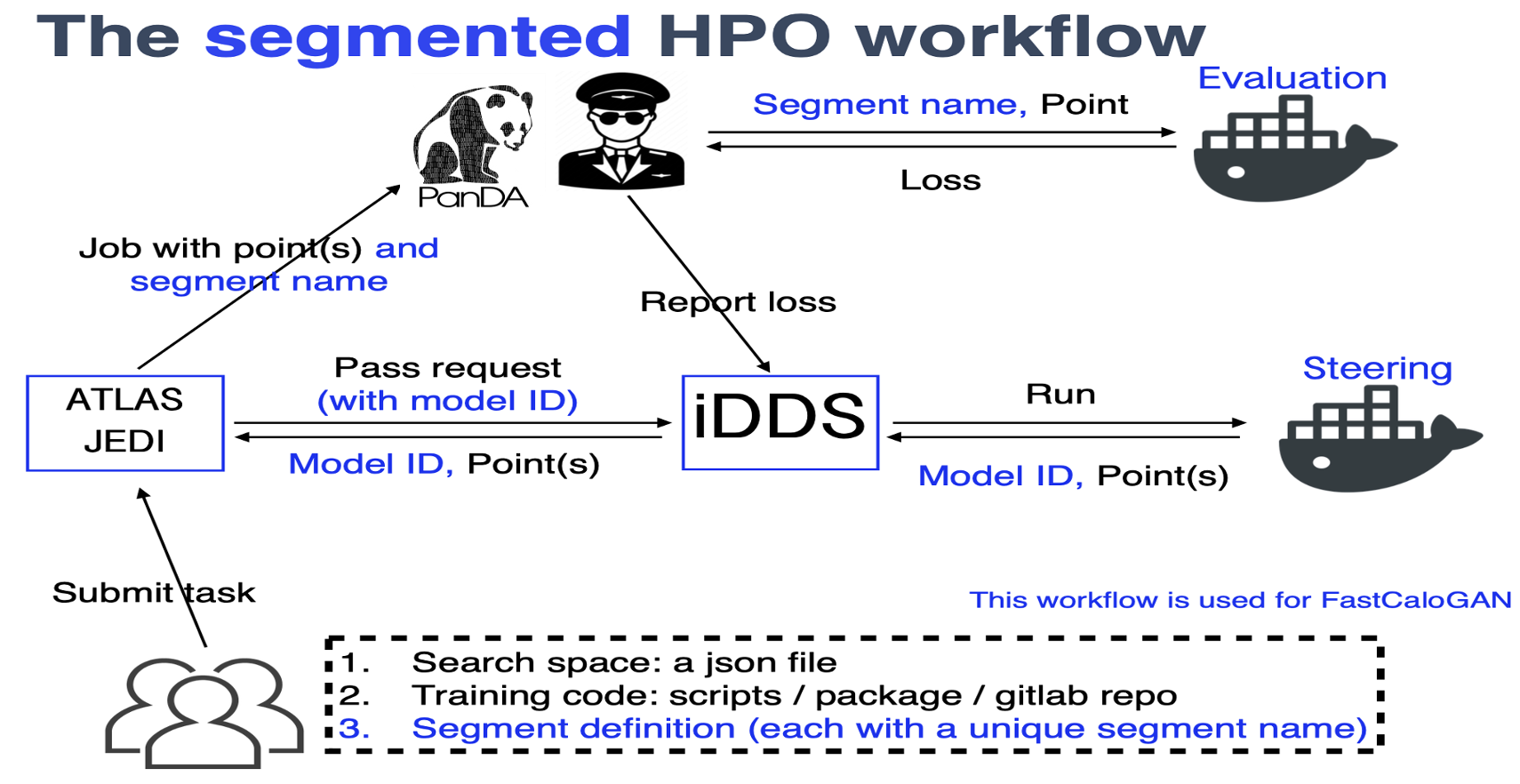} }%
    \caption{Architecture of the iDDS Hyperparameter Optimization service. }%
    \label{fig_hpo}
\end{figure}

A single iteration of HPO workflow consists of the following steps:

\begin{enumerate}
\item \textbf{Candidate sampling}: iDDS centrally explores the hyperparameter space using advanced search strategies such as Bayesian optimization~\cite{bayesian}, and chooses candidate parameter sets.
\item \textbf{Training with each candidate and evaluation}: The generated candidates are asynchronously dispatched to distributed computing sites via PanDA for model training and calculating performance metrics for each candidate.
\item \textbf{Search space refinement}: iDDS collects performance metrics to refine the search space and initiates the next iteration based on the refined space.
\end{enumerate}

This iterative process continues until the best-performing hyperparameters and corresponding trained models are identified. To further improve efficiency, iDDS supports segmented HPO, enabling the simultaneous optimization of multiple machine learning models. This approach increases throughput, reduces bias, and is well suited for ensemble learning and comparative model studies.


This HPO service is currently in production for ATLAS machine learning projects such as FastCaloGAN~\cite{fastsimulation}, where it has demonstrated strong scalability and performance. Although originally developed for ATLAS, the system is designed to be experiment-agnostic and can be readily extended to support HPO in a broad range of scientific and industrial ML applications.

\subsection{Active Learning}
\label{subsec:activelearning}

\begin{figure}[h]
\centering
\includegraphics[width=10cm]{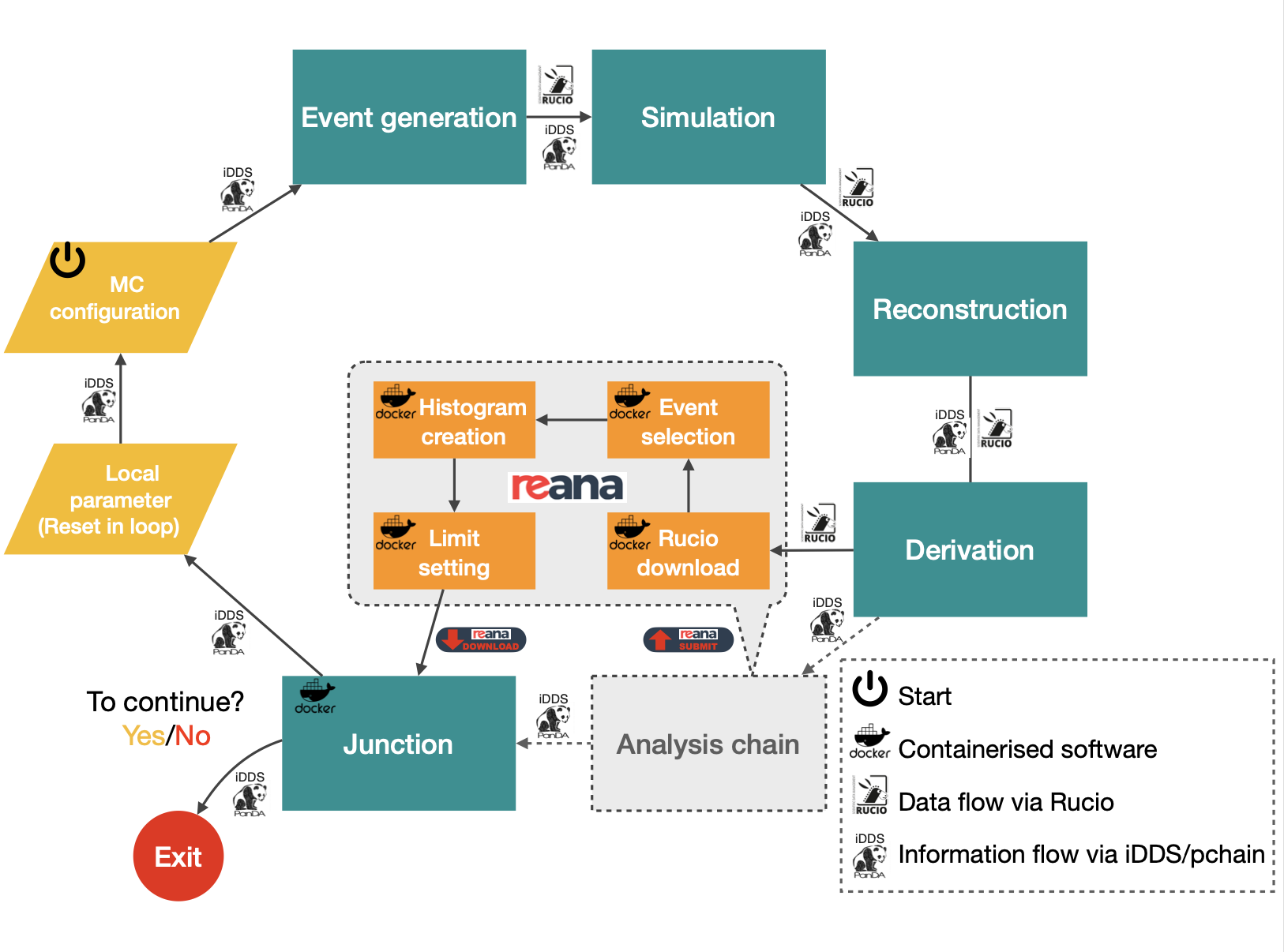}
\caption{A schematic view of the Active Learning workflow used in the $\textit{H} \rightarrow ZZ_d \rightarrow \textit{4l}$ analysis.}
\label{fig_al}
\end{figure}

Active Learning (AL)~\cite{activelearning,al,idds-activelearning} is an iterative machine learning technique that strategically refines the parameter space by incorporating feedback from previous results. Rather than processing the entire dataset in a single pass, AL selectively samples and labels new data points based on model uncertainty, enabling more efficient searches—especially in domains such as new physics analyses.

iDDS, in coordination with PanDA, enables fully automated, intelligent AL workflows by orchestrating iterations of training, uncertainty estimation, data selection, and retraining. Its conditional logic and dynamic branching capabilities make it particularly well-suited for AL use cases that demand tight feedback loops and scalable computation with minimal human intervention.

A representative AL workflow has been implemented for the optimized search of $\textit{H} \rightarrow ZZ_d \rightarrow \textit{4l}$~\cite{active_learning_dark}, as illustrated in Figure~\ref{fig_al}. The workflow comprises two primary stages: the production chain and the analysis chain. The \textit{production chain} starts from a template Monte Carlo (MC) configuration and proceeds through simulation and reconstruction to generate a Derived AOD (DAOD) sample for each physics parameter point. Once the DAOD is available, the \textit{analysis chain} is triggered, in which PanDA jobs execute a REANA workflow~\cite{reana} for data analysis and run Bayesian optimization to evaluate statistical significance, identify regions of interest (e.g., excesses between expected and observed limits), and propose new parameter points.

Throughout this process, iDDS manages all orchestration steps, ensuring coordination between components, data flow, and dynamic generation of new workloads—without human intervention. The workflow has successfully demonstrated AL-driven re-analysis capabilities and was published in an ATLAS Public Note~\cite{active_learning_dark}. A second AL workflow is currently under development for a generic Heavy Higgs $\rightarrow WW$ search.

\subsection{AI-assisted Detector Design at EIC (AID2E)}\label{subsec:aid2e}

The AI-assisted Detector Design for the Electron-Ion Collider (AID2E~\cite{aid2e_2024}) project showcases the effective use of iDDS in orchestrating iterative workflows involving geometry generation, simulation, reconstruction, and analysis. Metadata and performance metrics from previous runs are incorporated to guide the subsequent iterations, enabling efficient exploration of detector configurations.

\begin{figure}[h]
\centering
\includegraphics[height=6cm, keepaspectratio, width=12cm]{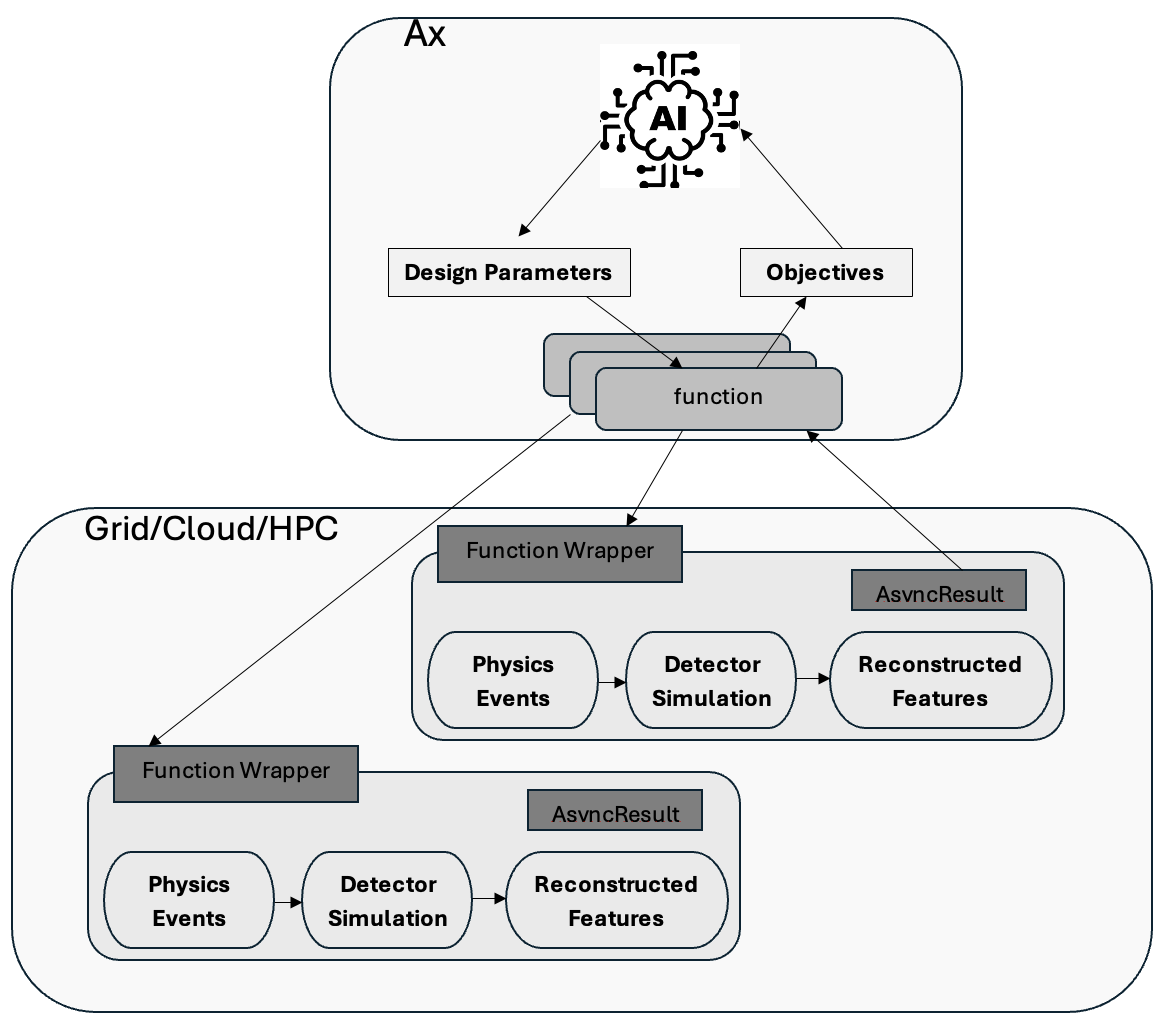}
\caption{AID2E: The Function-as-a-Task model maps local evaluation logic to distributed computing resources using a Python decorator that transforms a local function into a distributed task.}
\label{fig:aid2e_overall}
\end{figure}

Due to the complexity of AID2E, directly converting its workflow into distributed PanDA jobs is non-trivial and typically requires substantial restructuring. To simplify this process, we apply the iDDS Function-as-a-Task model, which streamlines the transformation of local workflows into distributed applications. Function-as-a-Task uses Python decorators to convert local functions in the AID2E code into distributed tasks composed of multiple concurrent jobs. With minimal code changes, users can transparently offload computations to PanDA-managed resources, enabling scalable parallel execution. Building on this capability, we have integrated Function-as-a-Task into AID2E. Each optimization cycle evaluates multiple detector configurations through simulation and assesses performance metrics such as resolution. These results guide the generation of new configurations in subsequent iterations, continuing until performance targets are met.
Throughout the process, iDDS ensures reproducibility and scalability by managing parameter sets, tracking metadata, and maintaining dependencies across iterations. This enables optimization algorithms to iteratively refine detector designs based on data-driven feedback.

By leveraging iDDS and PanDA, AID2E achieves scalable, high-throughput execution across diverse computing infrastructures—including Grid, Cloud, and HPC environments—significantly accelerating the detector design optimization process.

\section{Conclusion and Outlook}

iDDS provides a unified, scalable, and programmable platform for managing distributed workloads and data orchestration in large-scale scientific environments. The template-based and code-based workflow representations provide users with great flexibility to define workflows that accommodate dynamic runtime conditions and resource constraints. Its support for conditional logic, parameter passing, and result tracking makes it particularly well-suited for modern scientific workflows involving simulation, reconstruction, data analysis, and AI.

Integration with established infrastructures, such as PanDA and Rucio, allows iDDS to minimize the need for new resource configurations and leverage existing diverse resources. Its asynchronous result retrieval service, metadata handling, and execution tracking provide a reliable foundation for reproducible and efficient computation.

Looking ahead, iDDS will continue to evolve to support emerging computational paradigms. Ongoing developments aim to enhance the user experience, broaden support for interactive and serverless workflows, and integrate with cloud-native ecosystems.

As scientific computing becomes increasingly data-driven and iterative, iDDS is positioned to be a cornerstone of intelligent workflow orchestration, enabling new discoveries through automation, adaptability, and scalability.

\backmatter

\section*{Acknowledgments}

This work was done as part of the distributed computing research and development program within the ATLAS Collaboration. We thank our ATLAS colleagues for their support. In particular, we wish to acknowledge the contributions of the ATLAS Distributed Computing (ADC) team. Copyright 2024 CERN for the benefit of the ATLAS Collaboration. Reproduction of this article or parts of it is allowed as specified in the CC-BY-4.0 license. \\

\noindent This manuscript has been authored by employees of Brookhaven Science Associates, LLC under Contract No. DE-SC0012704 with the U.S. Department of Energy. The publisher by accepting the manuscript for publication acknowledges that the United States Government retains a nonexclusive, paid-up, irrevocable, worldwide license to publish or reproduce the published form of this manuscript, or allow others to do so, for United States Government purposes.

\noindent The iDDS project was initially supported by IRIS-HEP from 2020 to 2022 through the National Science Foundation under Cooperative Agreement OAC-1836650.



\bibliography{sn-bib}

\end{document}